\documentclass[gsifonts,twoside,12pt]{gsipaper}
\usepackage{a4wide,gsiindex,helvet}
\usepackage[english]{babel}
\usepackage{amsmath,amsfonts,amssymb,verbatim,graphicx,subfigure,float,epsfig,
psfrag,enumitem}
\begin{document}

\title{Wilson Loop-Loop Correlators in AdS/QCD}

\abstract{We calculate the expectation value of one circular Wilson loop and
the correlator of two concentric circular Wilson loops in AdS/QCD using the
modified $AdS_5$-metric given in Ref.~\cite{Pirner}. The confinement properties
of this metric in AdS/QCD are analyzed and compared with QCD and Nambu-Goto
theory in four dimensions.}

\author{J. Nian$^a$, H. J. Pirner$^{ab}$}

\address{$\phantom{!}^{a}$ Institute for Theoretical Physics, University of
Heidelberg, Germany\\$\phantom{!}^{b}$ Max Planck Institute for Nuclear Physics,
Heidelberg, Germany}

\maketitle
\thispagestyle{empty}

\section{Introduction}\label{introduction}
The $AdS/CFT$ correspondence has been developed by Maldacena \cite{Maldacena97},
who has calculated the heavy quark anti-quark potential with the aid of the
Wilson loop \cite{Maldacena}. This method does not provide a confining term in
the potential, i.e. the Wilson loop does not obey the area law for large areas,
since the $AdS_5$-metric has conformal symmetry in contrast to QCD. Many
suggestions have been made since then, to find a more QCD-like model. A
convenient way is to introduce a factor in the $AdS_5$-metric, which
breaks the conformal symmetry. In this way, the correct heavy quark anti-quark
potential with confinement can be reproduced (see e.g. \cite{Pirner}
\cite{Andreev}).

Loop-loop correlations have been an important tool to analyze non-perturbative
features of QCD. For long distances they can be used to investigate the lowest
lying glueball mass. In high energy scattering, light-like Wilson loops describe
the fast moving projectile and target hadron \cite{Shoshi:2002in}. In Euclidean
space, correlators of inclined Wilson loops can be used to describe the
S-Matrix for hadron-hadron scattering \cite{Shoshi:2002rd}. Before calculating
the loop-loop correlation function in these complicated circumstances, simpler
geometric configurations have to be explored. This is the goal of the present
work. We calculate the expectation value of one circular Wilson loop and the
correlator of two concentric circular Wilson loops using an improved confining
$AdS_5$-metric given in Ref.~\cite{Pirner} and the conformal $AdS_5$-metric in
five dimensions. The results can be related to the results of the heavy quark
anti-quark potential, and compared with the minimal area law.

\section{Vacuum Expectation Value of One Circular Wilson Loop 
$\langle\mathrm{W}\rangle$}\label{SectionOneloop}
The basic ideas to calculate the expectation values of one Wilson loop
$\langle\mathrm{W}\rangle$ and of two Wilson loops $\langle\mathrm{WW}\rangle$
are similar. Based on Ref.~\cite{Maldacena}, they are both equal to
$e^{-\mathrm{S}_{\mathrm{NG}}}$, where $\mathrm{S}_{\mathrm{NG}}$ is the
Euclidean five-dimensional Nambu-Goto action, which is proportional to the area
of the minimal surface stretched by the contours of the Wilson loops in the
$AdS_5$-space. The specific calculations of $\langle\mathrm{W}\rangle$
and $\langle\mathrm{WW}\rangle$ vary, because they require different boundary
conditions, i.e. the minimal surface is bounded either by one
or by two contours.

\begin{figure}[!ht]
  \begin{center}
  \epsfxsize 12cm
  \epsffile{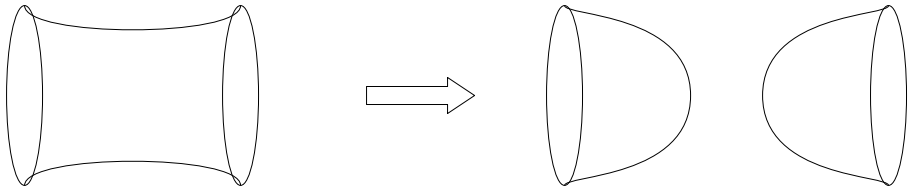}
  \end{center}
  \caption{\small Separation of the surface connecting two Wilson loops into two
disconnected surfaces}
\label{sketch}
\end{figure}

The minimal surface bounded by two concentric circular contours of the same
radius in four-dimensional flat space is a catenoid, with which we will compare
our final results from five-dimensional string theory. The results of
QCD-lattice calculations would serve as the main yard stick for our
phenomenology, but to our knowledge there are no simulations with circular
loops. Besides possible similarities between these approaches there are
new features in $AdS_5$-space. An example is following: When the separation
between the two contours gets larger, and above some critical value, the
surface splits into two parts, since the action from two separated surfaces
becomes smaller than the action of the connected surface (see
Fig.~\ref{sketch}).
\\
Therefore, at some critical distance, a phase transition is expected, which
is sometimes called ``Gross-Ooguri phase transition'' \cite{gross}. In order to
find this critical distance, we need both $\mathrm{S}_{\mathrm{NG}}$ for one
Wilson loop and $\mathrm{S}_{\mathrm{NG}}$ for two Wilson loops.

We first calculate the expectation value of one Wilson loop. This helps to
define our notations. The expectation value of one circular Wilson loop in
$AdS/CFT$ has been studied in Ref.~\cite{AdSoneloop} by using the special
conformal
transformation. Here we will reproduce the same result in a different way.

\begin{figure}[!ht]
  \begin{center}
  \epsfxsize 4.2cm
  \epsffile{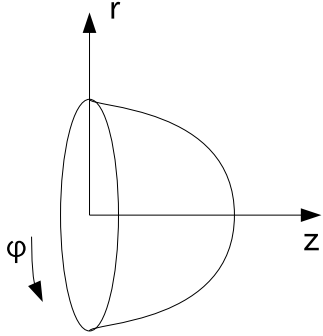}
  \end{center}
  \caption{\small Cylindrical coordinate system in $AdS_5$-space (the third
spatial coordinate $x$ is not plotted.)}
\label{oneloop}
\end{figure}

The Wilson loop lies in the $(x_1,x_2)$-plane defined by the polar coordinates
$r$ and $\varphi$. Due to the symmetry, cylindrical coordinates are appropriate.
In five-dimensional Euclidean space (c.f. Fig.~\ref{oneloop}) we have five
coordinates:
\begin{itemize}
  \item $t$ - time
  \item $z$ - the extra dimension
  \item $r,\varphi,x$ - three spatial cylindrical coordinates
\end{itemize}
Later, also the two Wilson loops will be localized in the $(x_1,x_2)$- or
$(r,\varphi)$-plane and separated along the $x=x_3$-direction. The
two-dimensional surface on top of the
Wilson loop can be parametrized by two parameters $\sigma$ and
$\tau$. For one circular Wilson loop, it is natural to identify the azimuthal
angle $\varphi$ with $\sigma$, and leave $\tau$ as an independent variable.
Since we are interested in a static solution, the time $t$ can be set to 0. 
Therefore, the five coordinates have the following form:
\begin{equation}
  t=0,\,\varphi=\sigma,\,r=r\left(\tau\right),\,x=x\left(\tau\right),\,
z=z\left(\tau\right).
\label{coordinates}
\end{equation}

The metric of the $AdS_5$-space is conformal in $AdS/CFT$. In non-conformal
$AdS/QCD$ theory, the calculation of the expectation value of a circular Wilson
loop differs from the conformal calculation by deforming the
conformal $AdS_5$-metric. As we argued in Section~\ref{introduction}, the
conformal $AdS_5$-metric generates a Coulombic heavy quark anti-quark potential
\cite{Maldacena}, but cannot yield a confining potential. Therefore in QCD, we
must break the conformal symmetry. We follow the choice made in
Ref.~\cite{Pirner}, where the warp factor similar to the running coupling was
advocated. The Euclidean modified $AdS_5$-metric becomes
\begin{equation} 
  ds^2 = \frac{L^2\cdot
h(z)}{z^2}\left(dt^2+dz^2+dx^2+dr^2+r^2d\varphi^2\right),
\label{nonAdSmetric}
\end{equation}
where 
\begin{equation}
  h(z) = \frac{\mathrm{log}(\frac{1}{\epsilon})}{\mathrm{log}
\left[\frac{1}{(\Lambda z)^2+\epsilon}\right]},
\label{factorh}
\end{equation} 
and
$$\epsilon\equiv\frac{l_s^2}{L^2} = 0.48,\quad \Lambda\,\equiv\,\frac{1}{L} =
264\,\mathrm{MeV}.$$
We should emphasize, that the conformal $AdS_5$-metric can be easily regained,
simply by setting $h(z)=1$. So in the following, we derive the equations of
motion with this modified $AdS_5$-metric, and treat the $AdS/CFT$ solution
calculated from the conformal $AdS_5$-metric as a special case with $h(z)=1$.

The form of $h(z)$, Eq.~(\ref{factorh}), is based on the idea that the fifth
coordinate of five-dimensional string theory is related to the energy
resolution by $1/z \approx E $. One may keep conformal symmetry at UV-scale when
$z\rightarrow 0$, but increase the warping of the metric in the infrared,
similar to the increase of the QCD running-coupling $\alpha_s(p=1/z)$. The
$\Lambda$-parameter of QCD becomes the parameter breaking the scale invariance.
The form of the modified metric from the above phenomenology can be made
consistent with the five-dimensional gravity by constructing a dilaton
potential, which will be no longer constant as in $AdS/CFT$. The dilaton
potential plays the dynamical role creating the modified metric. This crude
phenomenological metric, of course, does not reproduce the correct beta function
of QCD,
but in fact models can be constructed incorporating the QCD-running correctly 
c.f. Refs.~\cite{Kiritsis-1, Kiritsis-2, Gravity}.
This metric given by Eq.~(\ref{nonAdSmetric}), can reproduce the heavy quark
anti-quark potential with a Coulombic and a confining term in quite good
agreement with the Cornell potential \cite{Pirner}.

As mentioned before, the vacuum expectation value of one Wilson loop
$\langle\mathrm{W}\rangle$ can be calculated in the following way:
\begin{equation}
  \langle\mathrm{W\left(C_1\right)}\rangle = e^{-\mathrm{S}_{\mathrm{NG}}},
\label{basic}
\end{equation} 
where $\mathrm{S}_{\mathrm{NG}}$ denotes the Euclidean Nambu-Goto action
given by 
\begin{equation}
  \mathrm{S}_{\mathrm{NG}} = \frac{1}{2\pi l_s^2}\int d\sigma
    d\tau \sqrt{\mathrm{det}\,g_{\alpha\beta}} = \frac{1}{2\pi
    l_s^2}\int d\sigma d\tau
    \sqrt{\mathrm{det}\,G_{\mu\nu}\,\partial_{\alpha}x^{\mu}\,
\partial_{\beta}x^{\nu}}.
\label{NG}
\end{equation} 
In the above expression, $l_s$ is the string length, and $g_{\alpha\beta}$
stands for the induced metric on the two-dimensional surface. Then the
Nambu-Goto action, Eq.~(\ref{NG}), with the coordinates given by
Eq.~(\ref{coordinates}) has the following form:
\begin{eqnarray}
  \mathrm{S}_{\mathrm{NG}} & = & \frac{1}{2\pi
  l_s^2}\int_0^{2\pi}d\varphi\,\int d\tau\frac{L^2 h(z)
  r}{z^2}\sqrt{\left(z'\right)^2+\left(x'\right)^2+\left(r'\right)^2},\\
  {} & = & \frac{1}{\epsilon}\int d\tau\frac{h(z)\cdot
r}{z^2}\sqrt{\left(z'\right)^2+\left(x'\right)^2+\left(r'\right)^2}.
\label{nonAdSNG}
\end{eqnarray} 
where the prime ($'$) denotes a derivative with respect to $\tau$.

Variations of this action give the general form of equations of motion:
\begin{equation}
\left(\frac{h(z)\cdot r}{z^2}\frac{r'}{\sqrt{
\left(z'\right)^2+\left(x'\right)^2+\left(r'\right)^2}}
\right)'-\frac{h(z)}{z^2}\sqrt{
\left(z'\right)^2+\left(x'\right)^2+\left(r'\right)^2} = 0,
\label{nonAdSEOM-1}
\end{equation}
\begin{equation}
  \left(\frac{h(z)\cdot r}{z^2}\frac{z'}{\sqrt{
\left(z'\right)^2+\left(x'\right)^2+\left(r'\right)^2}}
\right)'-\frac{r\cdot\left(\dot{h}(z)z-2h(z)\right)}{z^3}\sqrt{
\left(z'\right)^2+\left(x'\right)^2+\left(r'\right)^2} = 0,
\label{nonAdSEOM-2}
\end{equation}
\begin{equation}
  \frac{h(z)\cdot r}{z^2}\frac{x'}{\sqrt{
\left(z'\right)^2+\left(x'\right)^2+\left(r'\right)^2}} = k,
\label{nonAdSEOM-3}
\end{equation} 
where the prime ($'$) and the dot ($\dot{\phantom{a}}$) stand for derivatives
with respect to $\tau$ and $z$ respectively, and $k$ is an integration
constant.

Let us first study the $AdS/CFT$ solution. In the conformal $AdS_5$-metric,
$h(z)=1$, the metric simplifies
\begin{equation}
  ds^2= \frac{L^2}{z^2}(dt^2+d \vec x^2 +dz^2).
\label{AdSmetric}
\end{equation}
Variations of the action with respect to $r$, $z$ and $x$ give three simpler
equations of motion, of which the third one is now
\begin{equation}
  \frac{r}{z^2}\frac{x'}{\sqrt{
\left(z'\right)^2+\left(x'\right)^2+\left(r'\right)^2}} = k.
\label{AdSEOM}
\end{equation}
If we choose $\tau=x$, then all primes become derivatives with respect to
$x$. Combining Eqs.~(\ref{nonAdSEOM-1}) and (\ref{nonAdSEOM-2}) for $h(z)=1$
with Eq.~(\ref{AdSEOM}), we obtain a system of three equations of motion:
\begin{equation}
  r''-\frac{r}{k^2 z^4} = 0,
\label{AdSEOM-1}
\end{equation}
\begin{equation}
  z''+\frac{2 r^2}{k^2 z^5} = 0,
\label{AdSEOM-2}
\end{equation}
\begin{equation}
  \left(z'\right)^2+\left(r'\right)^2+1-\frac{r^2}{k^2
z^4}\,=\,0.
\label{AdSEOM-3}
\end{equation}
These equations give the following important relation for the solution in the
conformal $AdS_5$-metric:
\begin{equation}
  \left(r^2+z^2\right)''+2 = 0.
\label{AdScrucial}
\end{equation}
The general solution of Eq.~(\ref{AdScrucial}) is
\begin{equation}
  r^2(x)+z^2(x) = -x^2+c_1\cdot x+c_2,
\end{equation}
where the constants $c_1$ and $c_2$  have to be fixed by the boundary
conditions. The geometry of the circular Wilson loop gives (c.f. 
Fig.~\ref{oneloop})
\begin{equation}
  r(z=0) = R.
\end{equation}
Since the Wilson loop surface does not propagate in the $x$-direction, both $r$
and $z$ have to be evaluated at $x=0$ in the solution. Therefore we obtain
\begin{equation}
  r^2+z^2 = R^2,
\label{oneloopcond-1}
\end{equation}
and consequently
\begin{equation}
  \frac{dr}{dz} = -\frac{z}{r}.
\label{oneloopcond-2}
\end{equation}

To calculate the action $\mathrm{S}_{\mathrm{NG}}$, it is advantageous to use
the $\tau=z$ gauge, since $\mathrm{S}_{\mathrm{NG}}$ then depends only on $z$
and $\dot{r}=\frac{d r}{d z}$ at $x=0$:
\begin{equation}
  \mathrm{S}_{\mathrm{NG}} = \frac{1}{\epsilon}\int
dz\frac{r}{z^2}\sqrt{1+\dot{r}^2}.
\label{NGoneloop-1}
\end{equation}
For a finite $\mathrm{S}_{\mathrm{NG}}$, we have to regulate the
UV-singularity at $z=0$, by a cutoff $z_0$. Then Eq.~(\ref{oneloopcond-2}) and
Eq.~(\ref{NGoneloop-1}) yield
\begin{equation}
\mathrm{S}_{\mathrm{NG}} = \frac{1}{\epsilon}\left(\frac{R}{z_0}-1\right) = 
\frac{L^2}{l_s^2}\left(\frac{R}{z_0}-1\right).
\end{equation} 
This result is the same as Eq.~(3.6) of Ref.~\cite{AdSoneloop}. After removing
the divergence arising from $z_0\rightarrow 0$, we obtain the regularized
$\mathrm{S}_{\mathrm{NG}}$:
\begin{eqnarray}
  \mathrm{S}_{\mathrm{NG,reg}} & = & -\mathrm{log}\langle
\mathrm{W}(C_1)\rangle\\
  {} & = & -\frac{1}{\epsilon} \,\,=\,\, -\frac{L^2}{l_s^2}.
\label{AdSoneloopresult}
\end{eqnarray} 
This is the result for one circular Wilson loop. For two non-interacting
circular Wilson loops, the total action is simply two times the one for one
circular Wilson loop. This value gives an upper limit for the five-dimensional
action of two Wilson loops. We emphasize that in the conformal $AdS_5$-metric 
the regularized Nambu-Goto action does not depend on the radius $R$ of the
Wilson loop, which differs from the result in the modified $AdS_5$-metric as we
will see.

Now we turn to the modified $AdS_5$-metric. Taking the constraint $x=0$ into
account, the Nambu-Goto action becomes
\begin{equation}
    \mathrm{S}_{\mathrm{NG}} = \frac{1}{\epsilon}\int dz\frac{h(z)\cdot
    r}{z^2}\sqrt{1+\dot{r}^2}.
\label{nonAdSNGoneloop}
\end{equation}
Then we obtain one independent equation of motion:
\begin{equation}
  \frac{h(z)}{z^2\sqrt{1+\dot{r}^2}}\left[-\frac{2r\dot{r}}{z}
+\frac{\dot{h}(z)r\dot{r}}{h(z)}-\frac{r\dot{r}^2\ddot{r}}{1+\dot{r}^2}+r\ddot{r
}-1\right] = 0.
\label{nonAdSoneloopEOM}
\end{equation}
The boundary condition to this equation of motion is
\begin{equation}
  r(z=0) = R.
\label{nonAdSoneloopBoundary}
\end{equation}

\begin{figure}[!ht]
  \begin{center}
  \epsfxsize 6cm
  \epsffile{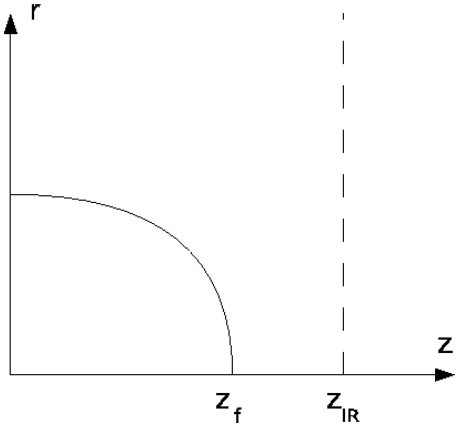}
  \end{center}
  \caption{\small The profile of the surface stretched by one Wilson loop in the
modified $AdS_5$-metric}
\label{nonAdSoneloop-1}
\end{figure}

The factor $h(z)$ implies an IR limit
$z_{\mathrm{IR}}=\sqrt{\frac{1-\epsilon}{\Lambda^2}}\approx 0.538\,\mathrm{fm}$.
The surface must end before this position and cannot go
beyond. Suppose the surface already closes at $z_f<z_{\mathrm{IR}}$. In this
case we have the natural second boundary condition besides
Eq.~(\ref{nonAdSoneloopBoundary}) at $z=z_f$:
\begin{equation}
  r(z=z_f) = 0,\label{nonAdSoneloopBoundary-1}
\end{equation} 
For an arbitrary $z_f$ with $0<z_f<z_{\mathrm{IR}}$, the first four terms in the
bracket of Eq.~(\ref{nonAdSoneloopEOM}) all vanish and only the summand $-1$
remains, which necessitates $\dot{r}(z)=\pm\infty$ to make the equation correct,
since $h(z_f)<\infty$. This infinitely steep fall of $r(z)$ at $z_f$ is shown
in Fig.~\ref{nonAdSoneloop-1}.
\\
For a given size $R$ of the Wilson loop, Eq.~(\ref{nonAdSoneloopEOM}) can be
solved by a shooting method. We vary the value of $z_f$, and start integrating
with an infinite derivative aiming at $r=R$ for $z=0$.

\begin{figure}[!htb]
  \begin{center}
  \epsfxsize 8.5cm \epsffile{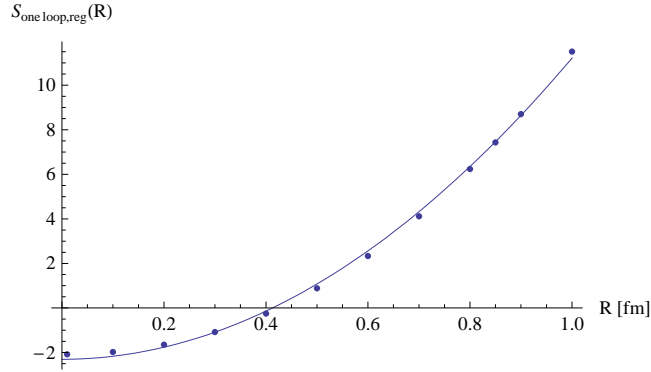}
  \end{center}
  \caption{\small Calculation in the modified $AdS_5$-metric of
the regularized action $\mathrm{S}_{\mathrm{NG,reg}}$ for circular Wilson loops 
with radii $R$}
\label{SRnonAdSoneloop}
\end{figure}

Once we have the result $r(z)$, we can calculate the Nambu-Goto action for one
circular Wilson loop in the modified $AdS_5$-metric using
Eq.~(\ref{nonAdSNGoneloop}), and also the corresponding
$\langle\mathrm{W}\rangle$ using Eq.~(\ref{basic}). Unlike the conformal case,
the result depends on the radius $R$ of the Wilson loop. In
Fig.~\ref{SRnonAdSoneloop}, the regularized $\mathrm{S}_{\mathrm{NG}}$ is
plotted for $R<1\,\mathrm{fm}$.
Using the results in this interval of $R$, one can fit the computed
$R$-dependence with the form
\begin{equation}
  \mathrm{S}_{\mathrm{one\,loop,reg}}(R)\,=-\frac{1}{c_1}+c_2\cdot \pi R^2.
\end{equation}
The values of $c_1$ and $c_2$ are
\begin{displaymath}
  c_1=0.43,\quad c_2=0.17\,\mathrm{GeV}^2,
\end{displaymath}
which are very close to $\epsilon$ and the string tension $\sigma$,
respectively. This result can be interpreted as a combination of the conformal
dependence at short distances (c.f. Eq.~(\ref{AdSoneloopresult})) given by the
dimensionless parameter $\epsilon=0.48$, and a confining behavior at large
distances given by the string tension $\sigma$. For the calculation of the heavy
quark anti-quark potential, i.e. a rectangular Wilson loop,
the string tension was estimated for AdS/QCD in Ref.~\cite{Pirner}
\begin{equation}
  \sigma = \left(-\frac{1}{4\pi} + \frac{27}{256\pi}
\frac{\Gamma(1/4)^2}{\Gamma(7/4)^2} \right) \frac{\Lambda^2}{\epsilon^2
\mathrm{log}(1/\epsilon)}.
\end{equation}
This value ($\sigma = 0.18\,\mathrm{GeV}^2$) is confirmed numerically for interquark 
distances $\Delta < 2\,\mathrm{fm}$ and agrees rather well with the result from the
circular loop above ($\sigma = 0.17\,\mathrm{GeV}^2$). For the heavy quark potential 
the analytical calculation from Refs.~\cite{Kinar, Kiritsis-1} gives the string 
tension analytically for asymptotic interquark distances $\Delta \rightarrow \infty$ 
with our metric as
\begin{equation}
  \sigma = \frac{h(z_*)}{2\pi\epsilon z_*^2} \approx 0.22\,\mathrm{GeV}^2,
\label{sigmatheo}
\end{equation}
where $z_*$ is given by
\begin{equation}
  \Lambda^2 z_*^2 + (\epsilon + \Lambda^2 z_*^2)\cdot \mathrm{log} (\epsilon + \Lambda^2 z_*^2) = 0.
\label{zstar}
\end{equation}
Numerically the fitting function yields $\sigma = 0.20\,\mathrm{GeV}^2$ even at 
$\Delta = 3.8\,\mathrm{fm}$, So the convergence is slow.

When the radius $R$ of the circular Wilson loop is very small, one can extract a
further term proportional to the square of the area in normal four-dimensional
space. As pointed out in Ref.~\cite{Banks} and Ref.~\cite{Rossi}, the gluon
condensate should be proportional to the coefficient of this
$\textrm{area}^2$-term for small area. Therefore we replot
Fig.~\ref{SRnonAdSoneloop} for $0.01\,\mathrm{fm}\leqslant R\leqslant
0.06\,\mathrm{fm}$.
\begin{figure}[!htb]
  \begin{center}
  \epsfxsize 8.5cm
  \epsffile{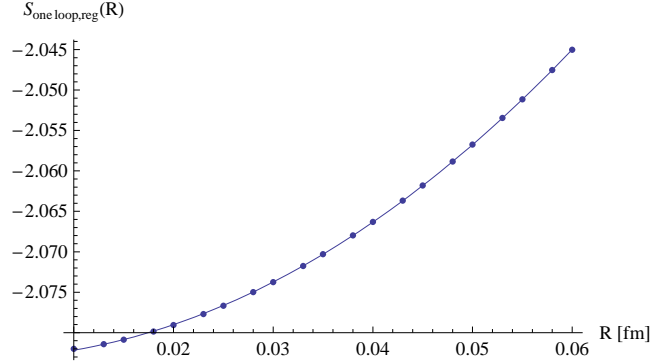}
  \end{center}
  \caption{\small Calculation in the modified $AdS_5$-metric of
the regularized action $\mathrm{S}_{\mathrm{NG,reg}}$ for extremely small
circular Wilson loops with radii $R$ }
\label{SRnonAdSoneloop-2}
\end{figure}
The curve in Fig.~\ref{SRnonAdSoneloop-2} can be fitted with a constant and two
more terms depending on the area and the squared area.
\begin{equation}
  \mathrm{S}_{\mathrm{one\,loop,reg}}(R) = -\frac{1}{d_1}+d_2 R^2+d_3 R^4,
\end{equation}
where
\begin{displaymath}
  d_1=0.48,\quad d_2=(10.51\pm 0.04)\frac{1}{\mathrm{fm}^2},\quad
d_3=(31.07\pm 9.79)\frac{1}{\mathrm{fm}^4}.
\end{displaymath}
Relating the extra $R^4$-term to the expectation value $\langle P\rangle$ of a
plaquette of size $a^2$ on the lattice (c.f. Ref.~\cite{Rossi}):
\begin{equation}
  \langle 1-P\rangle = \langle
\frac{\alpha_s}{\pi}\sum_{a,\mu\nu}G_{\mu\nu}^a G_{\mu\nu}^a \rangle
\frac{a^4\pi^2}{12 N_c}+\cdots
= \langle \frac{\alpha_s}{\pi}\sum_{a,\mu\nu}G_{\mu\nu}^a G_{\mu\nu}^a\rangle
\frac{\pi^4 R^4}{12 N_c}+\cdots,
\end{equation}
we calculate the the gluon condensate from the coefficient of the $R^4$-term:
\begin{displaymath}
  \langle \frac{\alpha_s}{\pi}\sum_{a,\mu\nu}G_{\mu\nu}^a
G_{\mu\nu}^a\rangle  = (0.017\pm 0.006)\,\mathrm{GeV}^4
\qquad\textrm{for $N_c$ = 3.}
\end{displaymath}

The gluon condensate $\langle\frac{\alpha_s}{\pi}G^2\rangle$ was introduced by
Shifman, Vainshtein and Zakharov within the framework of QCD spectral sum rules
\cite{Shifman}. It plays an important role in gluodynamics Recent values
from tau-decay and charmonium by different groups give
$(0.022\pm 0.004)\,\mathrm{GeV}^4$ \cite{Narison}. The non-vanishing
of the gluon condensate and its positive sign has been seen in the lattice by
the Pisa group \cite{Rossi,DiGiacomo} and more recently by P. E. Rakow
\cite{Rakow}. We remark that Ref.~\cite{Andreev-2} has used a
different modification of the conformal $AdS_5$-metric and obtained
\begin{displaymath}
  \langle \frac{\alpha_s}{\pi}\sum_{a,\mu\nu}G_{\mu\nu}^a
G_{\mu\nu}^a\rangle = (0.0100\pm 0.0023)\,\mathrm{GeV}^4\qquad\textrm{for
$N_c$ = 3.}
\end{displaymath}



\section{Correlator of Two Circular Wilson Loops
$\langle\mathrm{WW}\rangle$}\label{SectionTwoloops}
For simplicity, we study two concentric circular Wilson loops of the same 
radius, and assume that they have opposite orientations. Then a surface is
stretched in the $AdS_5$-space between the contours of these two Wilson loops.
The metric of the five-dimensional space influences the shape of the surface. In
the following, we give an explicit solution for the surface in the conformal
$AdS_5$-metric and in the modified $AdS_5$-metric. The problem of
calculating $\langle\mathrm{WW}\rangle$ in the conformal $AdS_5$-metric has been
studied by Zarembo \cite{Zarembo}, and there exists an analytic solution. We
will quickly summarize his work as extensively as it is necessary to see the
changes for the modified $AdS_5$-metric.

\begin{figure}[!ht]
  \begin{center}
  \epsfxsize 6cm
  \epsffile{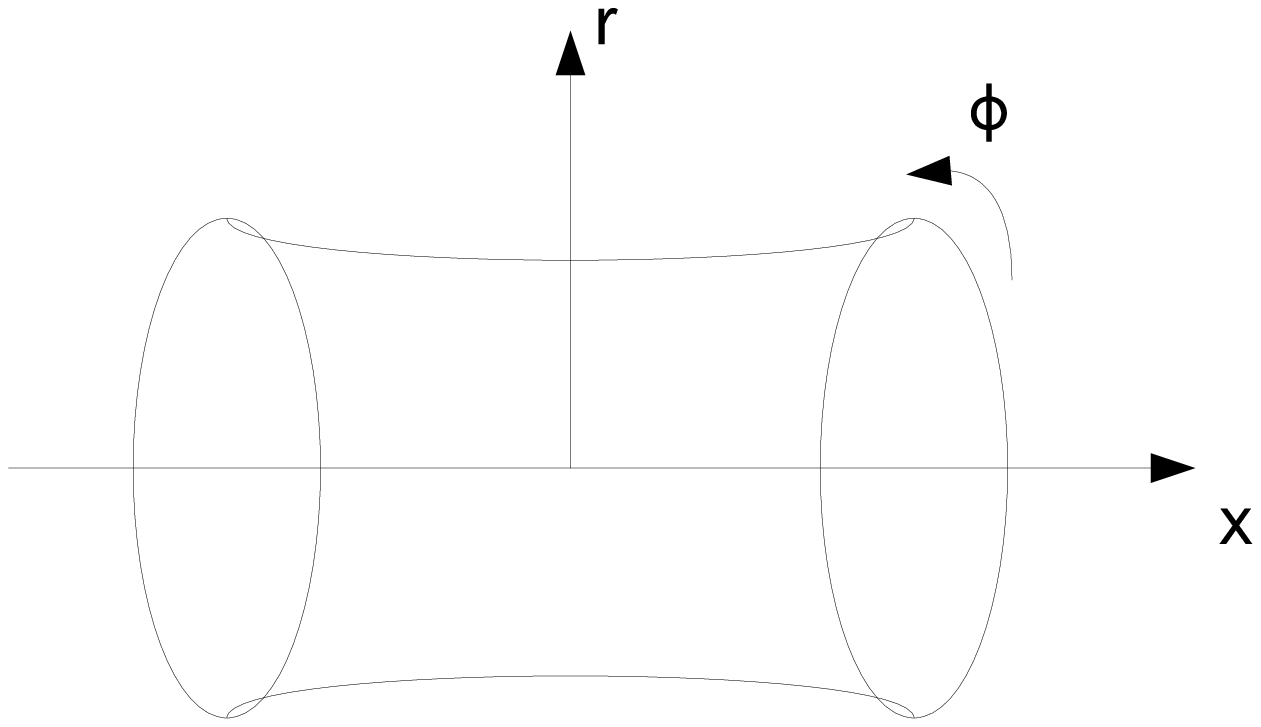}
  \end{center}
  \caption{\small Cylindrical coordinate system for two Wilson loops}
\label{twoloops}
\end{figure}

We can use the same cylindrical coordinate system in five-dimensional Euclidean
space as before. In four-dimensional space the two loops are arranged as shown
in Fig.~\ref{twoloops}.
The string surface connecting the contours of the Wilson loops can be
parametrized by two parameters $\sigma$ and $\tau$. We still choose
$\sigma=\varphi$ with $\tau$ variable and $t=0$. The Euclidean correlator
$\langle\mathrm{WW}\rangle$ is given by the Nambu-Goto action
$\mathrm{S}_{\mathrm{NG}}$ in $AdS_5$-space:
\begin{equation}
  \langle\mathrm{W\left(C_1\right)W\left(C_2\right)}\rangle =
e^{-\mathrm{S}_{\mathrm{NG}}}.
\label{NGtwoloops}
\end{equation}

First we solve the two-loop problem with the the conformal $AdS_5$-metric given
by Eq.~(\ref{AdSmetric}). For the choice $\tau=x$, the form of the equations
of motion (Eq.~(\ref{AdSEOM-1}) - Eq.~(\ref{AdSEOM-3})) remains the same
as in the case of one Wilson loop. Consequently, the relation
Eq.~(\ref{AdScrucial}) is also valid in the case of two Wilson loops.
The main difference comes from the boundary conditions, which for two
Wilson loops are
\begin{equation}
  r\left(-\frac{d}{2}\right) = r\left(\frac{d}{2}\right) = R,
\label{AdSBoundary-1}
\end{equation}
\begin{equation}
  z\left(-\frac{d}{2}\right) = z\left(\frac{d}{2}\right) = 0,
\label{AdSBoundary-2}
\end{equation}
where the two loops are located along the $x$-axis at $x=-d/2$ and $x=d/2$,
and $R$ is the common  radius of the loops. The crucial point to solve the
equations of motion analytically, is Eq.~(\ref{AdScrucial}):
\begin{displaymath}
  \left(r^2+z^2\right)''+2 = 0,
\end{displaymath}
which for the two boundary conditions, Eq.~(\ref{AdSBoundary-1}) and
Eq.~(\ref{AdSBoundary-2}), has the solution:
\begin{equation}
  r^2+z^2+x^2 = R^2+\frac{d^2}{4}\,\equiv\,a^2,
\end{equation}
where $a$ is constant for given $R$ and $d$. Due to this relation one can define
an angle $\theta$ by
\begin{equation}
  r\,\equiv\,\sqrt{a^2-x^2}\cdot\mathrm{cos}\theta,\quad
z\equiv\sqrt{a^2-x^2}\cdot\mathrm{sin}\theta,
\label{AdSSolEq-3}
\end{equation}
and arrive at a function $F(ka,\theta)$, which is already given by Zarembo
\cite{Zarembo}:
\begin{displaymath}
  F(ka,\theta) \equiv ka\int_0^{\theta}
\frac{d\phi\,\mathrm{sin}^2\phi}{\sqrt{\mathrm{cos}^2\phi-k^2 a^2
\mathrm{sin}^4\phi}}.
\end{displaymath}
It has to satisfy the following equations:
\begin{equation}
  F(ka,\theta)=\,\frac{1}{2}\mathrm{log}
\left(\frac{a+\frac{d}{2}}{a-\frac{d}{2}}\right)-\frac{1}{2}\mathrm{log}\left
(\frac{a-x}{a+x}\right),
\label{AdSSolEq-2}
\end{equation}
and
\begin{equation}
  F(ka,\theta_0) = \mathrm{log}\left(\frac{\sqrt{R^2+\frac{d^2}{4}}+
\frac{d}{2}}{R}\right),
\label{AdSSolEq-1}
\end{equation}
with
\begin{displaymath}
  \theta_0\equiv\theta(0)=\mathrm{arccos}\left(\frac{\sqrt{4 k^2
a^2+1}-1}{2ka}\right).
\end{displaymath}
When $R$ and $d$ are given, we can first solve Eq.~(\ref{AdSSolEq-1}) to
obtain the correct value of $ka$. Eq.~(\ref{AdSSolEq-2}) then gives an
implicit function $\theta(x)$. Inserting this function $\theta(x)$ into
Eq.~(\ref{AdSSolEq-3}), we find $r$ and $z$ as functions of $x$.

\begin{figure}[!ht]
  \begin{center}
  \epsfxsize 9cm
  \epsffile{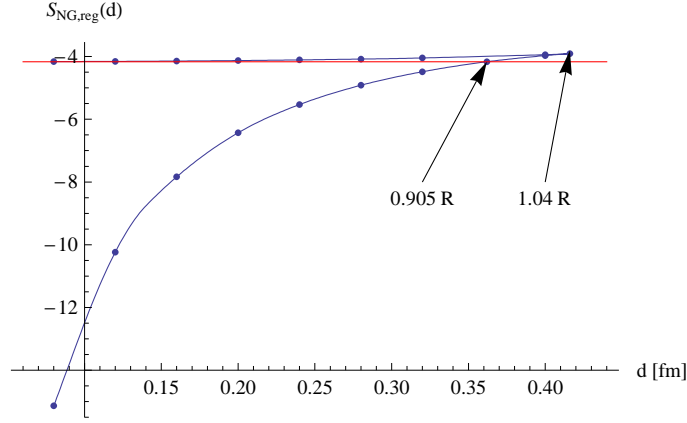}
  \end{center}
  \caption{\small The action $\mathrm{S}_{\mathrm{NG,reg}}(d)$ for two
    concentric circular Wilson loops of radii $R = 0.4\,\mathrm{fm}$
    calculated in the conformal $AdS_5$-metric: the lower curve with the ticks
    gives the solution with lower action, while the upper curve with ticks gives
the metastable solution. We also show the action for two disconnected Wilson
loops, which is a horizontal line.}
\label{AdS1+2}
\end{figure}

The final result for the Nambu-Goto action has to be regularized with $z_0$
as UV cutoff at the boundary, so that
\begin{equation}
  \mathrm{S}_{\mathrm{NG}}\,=\frac{L^2}{2 \pi
l_s^2}\left[\,\frac{4\pi R}{z_0}-4\pi\frac{\alpha}{\sqrt{\alpha-1}}
\cdot\int_0^{\frac{\pi}{2}}\frac{d\psi }{1 + \alpha 
\mathrm{sin}^2\psi+\sqrt{1+\alpha \mathrm{sin}^2\psi}}\right],
\end{equation}
where
\begin{equation}
  \alpha = \frac{1+2k^2a^2+\sqrt{1+4k^2a^2}}{2k^2a^2}.
\end{equation}
For small $d$, $\mathrm{S}_{\mathrm{NG}}$ has the asymptotic form: 
\begin{equation}
  \mathrm{S}_{\mathrm{NG}} = \frac{1}{\epsilon} \left(\frac{2
R}{z_0}-\frac{8\pi^3}{\Gamma^4\left(\frac{1}{4}\right)}\frac{R}{d}\right) =
\frac{2 R}{\epsilon z_0} - \frac{1}{\epsilon}
\frac{8\pi^3}{\Gamma^4\left(\frac{1}{4}\right)}\frac{R}{d}.
\label{AdSasympt}
\end{equation}
Since the UV-divergence is isolated, it can be removed to obtain the regularized
$\mathrm{S}_{\mathrm{NG}}$. In Fig.~\ref{AdS1+2} we show the regularized
action, $\mathrm{S}_{\mathrm{NG,reg}}$, as a function of the separation distance
$d$ of the two loops with size $R=0.4\,\mathrm{fm}$. For comparison, the action
for two non-interacting Wilson loops is also plotted for our choice of $\epsilon
= 0.48$:
\begin{equation}
  2 S_{\textrm{1-loop NG,reg}} = -2/\epsilon = -4.17.
\end{equation} 
The intersection of this curve with the curve describing the action of the
connected surface, defines the critical point at $d_c = 0.905R =
0.362\,\mathrm{fm}$, where it becomes advantageous to have two disconnected
surfaces. As it is well known for the $AdS/CFT$ solution, this happens earlier
than the limit for the existence of the solution at $d = 1.04R =
0.416\,\mathrm{fm}$.

Above the critical point $d_c = 0.905R = 0.362\,\mathrm{fm}$, two disconnected
surfaces give smaller Nambu-Goto action and are therefore more favorable. The
first derivative of the stable solution formed from the connected surface for
$d<d_c$ and the disconnected surface for $d>d_c$ has a discontinuity at
$d_c$. Therefore, one can say that a first-order phase transition, the so-called
``Gross-Ooguri phase transition'' \cite{gross}, should take place at this point.
We use the notations $S_{\mathrm{conn}}$ and $S_{\mathrm{disc}}$ for the
Nambu-Goto action of one connected surface and the Nambu-Goto action of two
disconnected surfaces, respectively. Then the correlation function of the Wilson
loops is given by\\
  for $d<d_c$:
  \begin{equation}
    \langle\mathrm{WW}\rangle-\langle\mathrm{W}\rangle
\langle\mathrm{W}\rangle = e^{-S_{\mathrm{conn}}}-e^{-2 S_{\mathrm{disc}}} \neq
0,
  \end{equation}
  for $d\geqslant d_c$:
  \begin{equation}
    \langle\mathrm{WW}\rangle-\langle\mathrm{W}\rangle
\langle\mathrm{W}\rangle = e^{-2
S_{\mathrm{disc}}}-e^{-2 S_{\mathrm{disc}}} = 0.
  \end{equation}
Considering the classical string surfaces connecting the two contours, we find
a nullification of the correlation function for $d\geqslant d_c$. Actually, when
$d\geqslant d_c$, the two Wilson loops are still weakly interacting through the
exchange of supergravitons, as discussed in Ref.~\cite{gross}. Therefore, the
correlation function will still have a non-vanishing value for $d\geqslant d_c$.

Now we want to calculate the correlator $\langle\mathrm{WW}\rangle$ in the
modified $AdS_5$-metric given by Eq.~(\ref{nonAdSmetric}). We use the 
same coordinate system as in the conformal $AdS_5$-metric, and the equations of
motion, Eqs.~(\ref{nonAdSEOM-1}) - (\ref{nonAdSEOM-3}), are completely the
same as in the case of one Wilson loop. To calculate
$\langle\mathrm{WW}\rangle$, $\tau=x$ is the right choice, then the
equations of motion, Eqs.~(\ref{nonAdSEOM-1}) - (\ref{nonAdSEOM-3}),
contain derivatives with respect to $x$ denoted by a prime ($'$) and derivatives
with respect to $z$ denoted by a dot ($\dot{\phantom{a}}$).

Eqs.~(\ref{nonAdSEOM-1}) and (\ref{nonAdSEOM-2}) contain all the dynamical
information on the system, while Eq.~(\ref{nonAdSEOM-3}) shows that the
system has a conserved quantity, like energy in a mechanical system. Therefore,
in order to solve $r(x)$ and $z(x)$ as functions of $x$, it is necessary to
solve the first two equations.

We may combine Eq.~(\ref{nonAdSEOM-1}), Eq.~(\ref{nonAdSEOM-2}) with
Eq.~(\ref{nonAdSEOM-3}), then the equations of motion in the modified
$AdS_5$-metric look similar to the ones in the conformal $AdS_5$-metric
(Eq.~(\ref{AdSEOM-1}) - Eq.~(\ref{AdSEOM-3})):
\begin{equation}
  r''-\frac{h^2(z)\cdot r}{k^2 z^4} = 0,
\label{nonAdSEOM-4}
\end{equation}
\begin{equation}
  z''-\frac{h(z)\cdot r^2\cdot\left(\dot{h}(z)z-2h(z)\right)}{k^2 z^5} = 0,
\label{nonAdSEOM-5}
\end{equation}
\begin{equation}
  \left(z'\right)^2+\left(r'\right)^2+1-\frac{h^2(z)\cdot r^2}{k^2 z^4} = 0.
\label{nonAdSEOM-6}
\end{equation}
We remark that the first two equations determine the
dynamics of $r(x)$ and $z(x)$ and the parameter $k$ in the third
equation has to be fixed in agreement with the boundary conditions.

The boundary conditions are given by the two contours as before:
\begin{equation}
  r\left(-\frac{d}{2}\right) = r\left(\frac{d}{2}\right) = R,
\label{nonAdSBoundary-1}
\end{equation}
\begin{equation}
  z\left(-\frac{d}{2}\right) = z\left(\frac{d}{2}\right) = z_0,
\label{nonAdSBoundary-2}
\end{equation}
where $d$ and $R$ stand for the separation of the two concentric contours and
their radii respectively, while $z_0$ gives the UV cutoff for $z$.

The difficulty of calculating $\langle\mathrm{WW}\rangle$ in the
modified $AdS_5$-metric originates from the fact that we do no longer have the
elegant relation Eq.~(\ref{AdScrucial}), instead we have now:
\begin{equation}
  \left(r^2+z^2\right)''+2-\frac{2h(z)\dot{h}(z)r^2}{k^2z^3} = 0.
\label{nonAdScrucial}
\end{equation}
Finding an analytic solution like in the conformal $AdS_5$-metric simply is not
possible in this case. Even a numerical solution to the system of equations is
problematic, since the boundary conditions, Eqs.~(\ref{nonAdSBoundary-1}) and
(\ref{nonAdSBoundary-2}), are given at two different positions. Analysis of
symmetry gives conditions for the derivatives $r'(0)=z'(0)=0$, and
conditions for the functions
$r\left(\pm\frac{d}{2}\right)=R,\,z\left(\pm\frac{d}{2}\right)=0$, but not at
the same position. We can convert this boundary-value problem into an
initial-value problem by studying the behavior of $r(x)$ and $z(x)$ near
$x\to\pm\frac{d}{2}$. Let us focus on the left boundary $x=-\frac{d}{2}$. Since
the equations of motion have translational symmetry in the $x$-direction, we
define a $v$-coordinate in such a way that the left boundary $x=-\frac{d}{2}$
becomes
$v=0$, i.e.
\begin{displaymath}
  v\equiv x+d/2.
\end{displaymath}
The asymptotic solutions $r_a(v)$ and $z_a(v)$ near $v=0$ have the following
forms:
\begin{displaymath}
  r_a(v)=R+a\cdot v^m,
\end{displaymath}
\begin{displaymath}
  z_a(v)=b\cdot v^n.
\end{displaymath}
The analysis of Eqs.~(\ref{nonAdSEOM-4}), (\ref{nonAdSEOM-6}) and
(\ref{nonAdScrucial}) yields
\begin{displaymath}
  m=\frac{2}{3},\quad n=\frac{1}{3}.
\end{displaymath}
Practically, we have chosen the following series as an asymptotic solution near
the boundary:
\begin{equation}
  r_a(v)=R+a_1\cdot v^{\frac{2}{3}}+a_2\cdot v^{\frac{4}{3}}+a_3\cdot
v^2+a_4\cdot v^{\frac{8}{3}}
\end{equation}
\begin{equation}
  z_a(v)=b_1\cdot v^{\frac{1}{3}}+b_2\cdot v+b_3\cdot
v^{\frac{5}{3}}+b_4\cdot v^{\frac{7}{3}}
\end{equation}
We insert these solutions with undetermined coefficients into
Eq.~(\ref{nonAdSEOM-6}) and Eq.~(\ref{nonAdScrucial}), and perform the
power-series expansion. Since the coefficients belonging to each power of $v$
must vanish, we have enough equations to determine the parameters in the
asymptotic solution for given values of $k$ and $R$.

Numerically, a small cutoff $v_0$ is applied, then $r(v_0)$, $z(v_0)$, $r'(v_0)$
and $z'(v_0)$ can be calculated directly and used as initial conditions for the
system of equations consisting of Eq.~(\ref{nonAdSEOM-4}) and
Eq.~(\ref{nonAdSEOM-5}). Unlike the procedure in the conformal case, we do
not prescribe the value of $d$. For a fixed value of $R$, we give an arbitrary
value $k>0$, then calculate $r(v)$, $z(v)$ and consequently the Nambu-Goto
action $\mathrm{S}_{\mathrm{NG}}$ for this value of $k$. By changing the value
of $k$ we obtain the Nambu-Goto action as a function of $k$, denoted by
$\mathrm{S}_{\mathrm{NG}}(k)$. For a given $k$, we search the position $v$,
where $r'(v)=z'(v)=0$. This $v$ corresponds to the mid-point ($v=\frac{d}{2}$)
between the two loops. This determines $d$, also as a function of $k$.
Combining $\mathrm{S}_{\mathrm{NG}}(k)$ with $d(k)$, we obtain
$\mathrm{S}_{\mathrm{NG}}(d)$.

\begin{figure}[!ht]
  \begin{center}
  \epsfxsize 8cm
  \epsffile{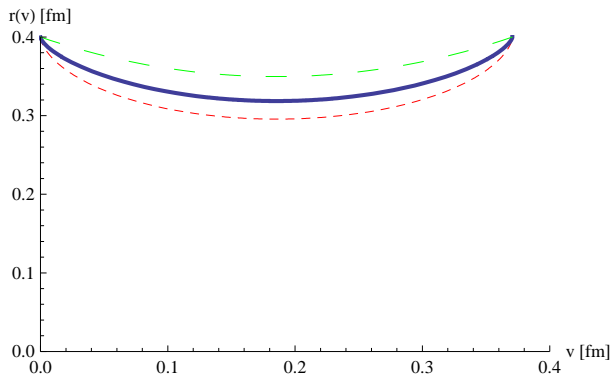}
  \end{center}
  \caption{\small The profile $r(v)$ of the surface connecting two Wilson loops
with $R=0.4\,\mathrm{fm}$ at a distance $d=0.37\,\mathrm{fm}$: The $v$-axis
shows the distance from the left loop located at $v=0$ or $x=-\frac{d}{2}$. The
continuous line is the result from the modified $AdS_5$-metric in $AdS/QCD$, the
short dashed line is the result from the conformal $AdS_5$-metric in $AdS/CFT$,
while the long dashed line gives the profile of the catenoid from the
four-dimensional Nambu-Goto action.}
\label{0.4solution-r}
\end{figure}

\begin{figure}[!ht]
  \begin{center}
  \epsfxsize 8cm
  \epsffile{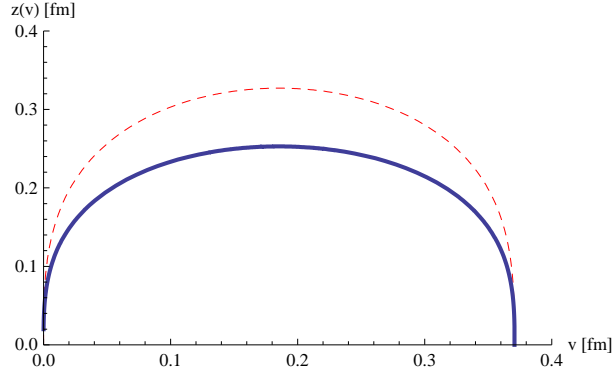}
  \end{center}
  \caption{\small The bulk coordinate $z(v)$ in the fifth dimension for the
surface connecting two loops with $R=0.4\,\mathrm{fm}$ and $d=0.37\,\mathrm{fm}$
as a function of the distance from the left loop located at $v=0$ or
$x=-\frac{d}{2}$: The continuous line is the result from the modified
$AdS_5$-metric in $AdS/QCD$, while the short dashed line is the result from the
conformal $AdS_5$-metric in $AdS/CFT$.}
\label{0.4solution-z}
\end{figure}

In Fig.~\ref{0.4solution-r} and Fig.~\ref{0.4solution-z}, we show the explicit
solutions $r(v)$ and $z(v)$ from the numerical calculation of two Wilson loops
of size $R = 0.4\,\mathrm{fm}$ in the conformal and the modified
$AdS_5$-metrics. One sees that the modified $AdS_5$-metric leads to solutions
different from the ones in the conformal $AdS_5$-metric. The differences in the
surface profile $r(v)$ and the bulk coordinate $z(v)$ do not look very large.
The modified metric with its wall at $z_{\mathrm{IR}}\approx 0.538\,\mathrm{fm}$
lets the surface dive less far into the fifth dimension than the conformal
$AdS_5$-metric.

The Nambu-Goto action is given by
\begin{eqnarray}
  \mathrm{S}_{\mathrm{NG}} & = & \frac{2}{\epsilon}\int_0^{\frac{d}{2}} dv
\frac{h(z)\cdot r}{z^2}\sqrt{1+\left(z'\right)^2+\left(r'\right)^2}\\
  {} & = & \frac{2}{\epsilon}\int_0^{v_0} dv \frac{h\left(z_a\right)\cdot
r_a}{z_a^2}\sqrt{1+\left(z_a'\right)^2+\left(r_a'\right)^2}\\
  {} & {} & +\frac{2}{\epsilon}
\int_{v_0}^{\frac{d}{2}} dv\frac{h\left(z_n\right)\cdot
r_n}{z_n^2}\sqrt{1+\left(z_n'\right)^2+\left(r_n'\right)^2},
\end{eqnarray}
where we use the asymptotic solutions $(r_a,z_a)$ for small $v$ and numerical
solutions $(r_n,z_n)$ for large $v$. In the last expression, the first integral
is divergent at $v=0$. But, as we have the explicit forms of the asymptotic
solutions $r_a(v)$ and $z_a(v)$, we can expand the first integrand into power
series near $v=0$, and remove the divergent terms. To compensate this removal,
we add the antiderivative of the divergent terms at $v=v_0$. In this way,
we obtain the regularized value of $\mathrm{S}_{\mathrm{NG}}$.

\begin{figure}[!ht]
  \begin{center}
  \epsfxsize 9cm
  \epsffile{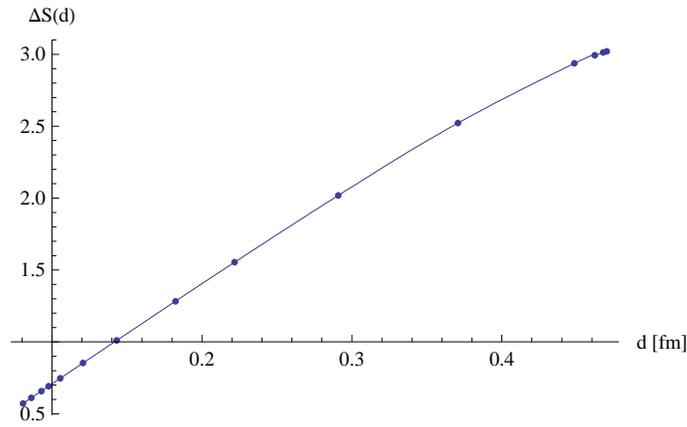}
  \end{center}
  \caption{\small $\Delta S(d)$ for
$R=0.4\,\mathrm{fm}$ in the modified $AdS_5$-metric}
\label{SLnonAdSsub}
\end{figure}

Using the method described above, we calculate the action
$\mathrm{S}_{\mathrm{NG}}$ of one continuous surface connecting the two
contours for a given value $R=0.4\,\mathrm{fm}$, and plot it together with the
result of two disconnected surfaces of the same radius given by
Eq.~(\ref{nonAdSNGoneloop}).
\begin{figure}[!ht]
  \begin{center}
  \epsfxsize 8cm
  \epsffile{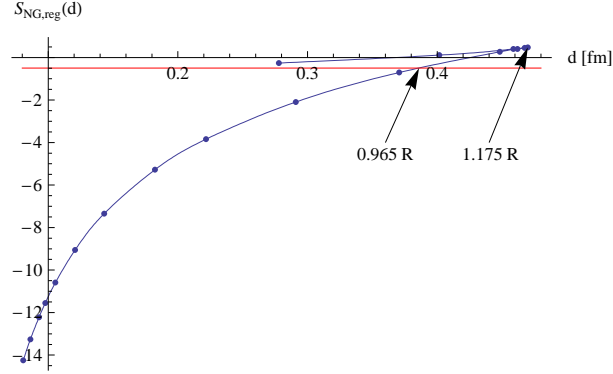}
  \end{center}
  \caption{\small $\mathrm{S}_{\mathrm{NG,reg}}(d)$ for $R=0.4\,\mathrm{fm}$
in the modified $AdS_5$-metric}
\label{nonAdS1+2}
\end{figure}
Fig.~\ref{nonAdS1+2} shows several new phenomena:
For $AdS/QCD$ the continuous surface can extend until a maximal value of
$d = 1.175R$, which exceeds the limit $d = 1.04R$ from $AdS/CFT$ by $20\%$. In
the $AdS/QCD$ case, above the critical point $d_c = 0.965 R$ two
separate surfaces become advantageous compared with one surface. This value of
$d_c$ is higher than in the $AdS/CFT$ case, where $d_c = 0.905R$. In the
conformal $AdS_5$-metric (see Fig.~\ref{AdS1+2}), we observe that the action
$\mathrm{S}_{\mathrm{NG}}$ behaves like $\frac{R}{d}$ in agreement with the
scale-free model, while the action in the modified $AdS_5$-metric
has an additional linear contribution towards the critical point $d_c$ (see
Fig.~\ref{nonAdS1+2}).
Since the warp factor $h(z) \rightarrow 1$ for $z \rightarrow 0$, we may assume
a form $\mathrm{S}_{\mathrm{NG}} = -\frac{a}{d}+b\cdot d$ for the $AdS/QCD$
action. In Fig.~\ref{SLnonAdSsub},
\begin{equation}
  \Delta S(d)\equiv \mathrm{S}_{\mathrm{NG,reg}}(d)-\left(-\frac{a}{d}\right) =
\mathrm{S}_{\mathrm{NG,reg}}(d)-\left(-\frac{1}{\epsilon}
\frac{8\pi^3}{\Gamma^4\left(\frac{ 1}{4}\right)}\frac{R}{d}\right)
\label{subtractedaction}
\end{equation}
is plotted, where the $\frac{1}{d}$-contribution given by the second term of
Eq.~(\ref{AdSasympt}) has been subtracted.
\\
The plot of Fig.~\ref{SLnonAdSsub} justifies our assumption. But we
must point out, that such a form is not good for very small values of $R$, e.g.
for $R\leqslant 0.2\,\mathrm{fm}$.

\begin{figure}[!ht]
  \begin{center}
  \epsfxsize 8.5cm
  \epsffile{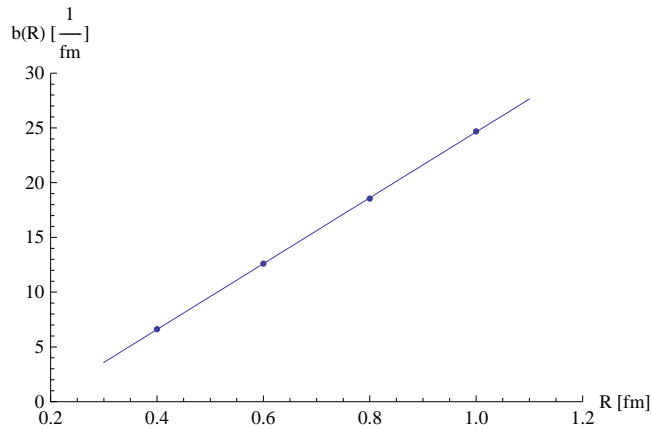}
  \end{center}
  \caption{\small The dependence of $b$ on $R$ in the modified $AdS_5$-metric}
\label{bR}
\end{figure}

Now let us turn to the discussion of the $R$-dependence of
$\mathrm{S}_{\mathrm{NG}}$. In the form $-\frac{a}{d}+b\cdot d$, the dependence
of $a$ on $R$ is given explicitly by Eq.~(\ref{AdSasympt}), so we only need to
consider how the slope $b$ depends on $R$. In Fig.~\ref{bR} we show the slope
$b$ for different values of $R$ in the modified $AdS_5$-metric. A
linear function with an off-set can describe the behavior very well:
\begin{equation}
  b(R) = \left(30.07\frac{1}{\mathrm{fm}^2}\right)\cdot(R-0.18\,\mathrm{fm}
).
\label{empirical-1}
\end{equation}
\\
Combining Eq.~(\ref{empirical-1}) with the $\frac{1}{d}$-like part given by
Eq.~(\ref{AdSasympt}), we propose an empirical formula, which is valid for
$0.4\,\mathrm{fm}\leqslant R\leqslant 1.0\,\mathrm{fm}$:
\begin{equation}
  \mathrm{S}_{\mathrm{NG}}(d,R)=-\frac{1}{\epsilon}\frac{8\pi^3}{
\Gamma^4\left(\frac{1}{4}\right)}\frac{R}{d}+\left(30.07\frac{1}{\mathrm{fm}^2}
\right)\cdot(R-0.18\,\mathrm{fm})d.
\label{empirical-2}
\end{equation}

\begin{figure}[!ht]
  \begin{center}
  \epsfxsize 8.5cm
  \epsffile{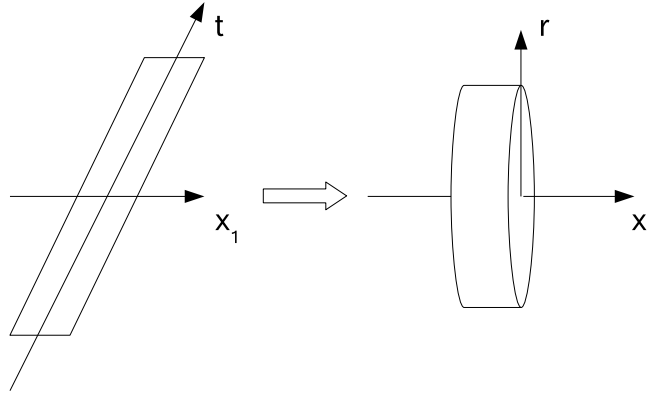}
  \end{center}
  \caption{\small The similarity of the surface bounded by the rectangular
contour and the surface formed by two concentric circles}
  \label{sketch-2}
\end{figure}

The physical interpretation of this formula is in accordance with the heavy
quark anti-quark potential calculated in Ref.~\cite{Pirner}. It
is well known \cite{Billoire, Eichten}, that the expectation value of a
rectangular Wilson loop of width $d$ and length $T\gg d$ in Euclidean space
obeys:
\begin{equation}
  \langle \mathrm{W}\rangle\,\simeq\,e^{-V(d)\cdot T},\quad T\gg d.
\label{wellknown}
\end{equation}
As the Wilson loop describes the phase factor of two infinitely heavy static
quarks, $V(d)$ is just the heavy quark anti-quark potential. Imagine that
we start with a rectangular contour with one side much larger than the other,
then we bend the long side into a circle. The correlator of two loops becomes
equivalent to the expectation value of one rectangular contour, where the short
side corresponds to the separation of the two circles, and the long side
corresponds to the circumference of the circular loop, i.e. $T=2 \pi R$ (see
Fig.~\ref{sketch-2}).
\\
Therefore, we expect for $2\pi R\gg d$:
\begin{eqnarray}
  \langle\mathrm{WW}\rangle & = & e^{-\mathrm{S}_{\mathrm{NG}}(d,R)}\\ 
  {}  & = & e^{-V_{\mathrm{Q\bar{Q}}}(d)\cdot 2\pi R}.
\label{wellknown-2}
\end{eqnarray}
Using the result for $V_{\mathrm{Q\bar{Q}}}(d)$ calculated in Ref.~\cite{Pirner}
with $\epsilon = 0.48$ and string tension $\sigma = 0.183\,\mathrm{GeV}^2$, we
obtain for $2\pi R\gg d$:
\begin{eqnarray}
  \mathrm{S}_{\mathrm{NG}}(d,R) & = &
\left[-2\left(\frac{\Gamma(3/4)}{\Gamma(1/4)}\right)^2\frac{1}{\epsilon
d}+\sigma d\right]\cdot 2\pi R,\\
  {} & = &
-\frac{1}{\epsilon}\frac{8\pi^3}{\Gamma^4\left(\frac{1}{4}\right)}\frac{R}{d}
+\left(29.63\frac{1}{\mathrm{fm}^2}\right)\cdot R\cdot d,
\end{eqnarray}
which is in good agreement with Eq.~(\ref{empirical-2}) except for $R\leqslant
0.2\,\mathrm{fm}$.

\begin{figure}[!ht]
  \begin{center}
  \epsfxsize 8cm
  \epsffile{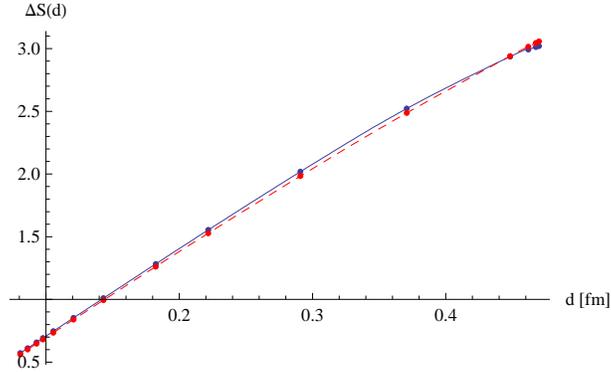}
  \end{center}
  \caption{\small Comparison of the five-dimensional $\Delta S(d)$ (continuous
line) with the catenoid solution (dashed line) in four dimensions using
$\sigma=0.108\,\mathrm{GeV}^2$}
  \label{catenoidfit}
\end{figure}

We can also compare our result with the minimal surface stretched by two
concentric circular contours in four dimensions. The area calculated 
from the Nambu-Goto action in four dimensions is given by the catenoid. Let us
briefly summarize the relevant results of the catenoid. If two circles of the
radius $R$ are located perpendicular to the $x$-axis at $x=-\frac{d}{2}$ and
$x=\frac{d}{2}$, then the profile of the catenoid
\begin{equation}
  r(x)=R_m\cdot \mathrm{cosh}\left(\frac{x}{R_m}\right)
\label{catenoid-1}
\end{equation}
 is determined by the radius $R_m$ at $x=0$, obeying the condition at
$x=\pm\frac{d}{2}$
\begin{equation}
  R=R_m\cdot \mathrm{cosh}\left(\frac{d}{2 R_m}\right).
\end{equation}
The area of the surface is given by
\begin{equation}
  A=\pi R_m d+\pi R_m^2 \mathrm{sinh}\left(\frac{d}{R_m}\right).
\label{catenoid-2}
\end{equation}
In Fig.~\ref{catenoidfit}, we compare the catenoid solution with the subtracted
Nambu-Goto action $\Delta S(d)$ from Eq.~(\ref{subtractedaction}) for $R =
0.4\,\mathrm{fm}$.
\\
The values of the string tension $\sigma=\Delta S(d)/A$ determined from the
area of the catenoid are listed in the following table:
\begin{table}[!ht]
\begin{center}
\begin{tabular}{|c||c|c|c|c|}
\hline
$R$ [$\mathrm{fm}$] & 0.4 & 0.6 & 0.8 & 1.0\\
\hline
$\sigma$ [$\mathrm{GeV}^2$] & 0.108 & 0.137 & 0.151 & 0.163\\
\hline
\end{tabular}
\end{center}
\end{table}\\
We find that the larger the radius of the contour is, the closer the
determined value of the string tension lies to the expected one, which is
$0.183\,\mathrm{GeV}^2$.


\section{Summary}
We have calculated the expectation value of a single circular Wilson loop in the
conformal $AdS_5$-metric and in the modified $AdS_5$-metric. We
indicated a different way from Ref.~\cite{AdSoneloop} to solve the one-loop
problem. In the modified $AdS_5$-metric we find the area law. For very
small loops, a term in the action proportional to the area squared can be
extracted and related to the gluon condensate. The numerical value for the gluon
condensate obtained from the parametrization \cite{Pirner} of the modified
$AdS_5$-metric is in good agreement with the phenomenological
\cite{Shifman,Narison} and the lattice values \cite{Rossi,DiGiacomo,Rakow}.

In Section~\ref{SectionTwoloops}, we have compared the correlator of two Wilson
loops defined for two concentric circular contours with opposite orientations in
conformal $AdS/CFT$ \cite{Zarembo} with the one calculated using the modified
$AdS_5$-metric. The results show that the modified $AdS_5$-metric produces
confinement with nearly the same string tension for the rectangular and circular
Wilson loops and for two circular Wilson loops. It is very important for further
studies of loop-loop correlators to have confinement and the short distance
Coulombic behavior. In previous work on loop-loop correlators
\cite{Shoshi:2002in,Shoshi:2002rd}, these two features had to be added by hand,
whereas here they follow from one action.


The general question can be asked whether the running of the 
QCD-coupling can be included in the five-dimensional geometrical picture of supergravity.
This necessitates a longer discussion for which we refer to the literature 
Refs.~\cite{Kiritsis-1, Kiritsis-2} and forthcoming work Ref.~\cite{Gravity}. In the large $N_c$-limit one
can construct a dilaton potential which is consistent with the QCD-running
coupling and leads to confinement. In the paper here the simple modified
metric with $h(z)$ has led to a consistent phenomenology for one and two Wilson loops.
The confining property of the metric is related to the strong increase
of the function $h(z)$, but not necessarily to the Landau-pole of 
this simple guessed metric. In fact for infinite separation of
quarks, the bulk coordinate stops sampling the metric before the Landau pole
cf. Eq.~(\ref{zstar}). Further phenomenological work on the improved holographic Yang Mills
theory  with a dilaton at finite temperature can be found most recently in 
Refs.~\cite{Kiritsis-3, Kajantie, Nian}.
  


\bibliographystyle{h-physrev3}
\bibliography{loopstemp}

\end{document}